\documentclass[
aps,%
12pt,%
final,%
notitlepage,%
oneside,%
onecolumn,%
nobibnotes,%
nofootinbib,% 
superscriptaddress,%
noshowpacs,%
centertags]%
{revtex4}

\begin{document}

\title{On the Limitations of Karmarkar's Condition in Static, Conformally Flat Spacetimes}

\author{\firstname{Samstuti}~\surname{Chanda}}
\email{schanda93.dta@gmail.com}
\affiliation{IUCAA Centre for Astronomy Research and Development, Department of Physics, Cooch Behar Panchanan Barma University, Cooch Behar, 736101, West Bengal, India}
\author{\firstname{Ranjan}~\surname{Sharma}}
\email{rsharma@associates.iucaa.in}
\affiliation{IUCAA Centre for Astronomy Research and Development, Department of Physics, Cooch Behar Panchanan Barma University, Cooch Behar, 736101, West Bengal, India}

\author{\firstname{Sunil}~\surname{Maharaj}}
\email{maharaj@ukzn.ac.za}
\affiliation{Astrophysics Research Centre, School of Mathematics, Statistics and Computer Science, University of KwaZulu-Natal, Private Bag 54001, Durban, 4000, South Africa}

\begin{abstract}
For a static and spherically symmetric spacetime, we investigate the class of exact solutions that arise when two fundamental geometric constraints are imposed simultaneously: the Karmarkar's condition and the vanishing of the Weyl tensor. These conditions restrict the curvature in such a way that the spacetime becomes conformally flat and belongs to the family of embedding class-I solutions. Even though the subsequent solutions namely, the Schwarzschild interior solution and the de Sitter solution are well known, the novelty of our presentation is that these solutions are shown to be a direct consequence of the imposed geometric constraints. The physical matter composition becomes highly constrained by the associated geometry under such conditions. The Schwarzschild interior solution describes the spacetime of an incompressible fluid sphere while the de Sitter solution corresponds to a vacuum energy dominated configuration. Interestingly, pressure anisotropy as well as `complexity factor' vanish identically once the Karmarkar's condition and the conformal flatness conditions are applied simultaneously. As these two geometric constraints alone are sufficient to determine the background spacetime uniquely, Karmarkar's condition might not be a suitable method for the development of realistic stellar models in a conformally flat spacetime unless one invokes other factors into consideration such as time-dependent metric potentials.
\end{abstract}
\maketitle

	\section{Introduction}\label{sec1}
	To model a relativistic compact star, one solves the Einstein field equations $G_{ab} = 8\pi G T_{ab}$, which are a highly nonlinear system of differential equations and, hence, difficult to solve unless some simplifying techniques are adopted. To develop a compact stellar model, either the left-hand side of the Einstein field equations, i.e., the geometry is specified by assuming a particular metric potential or the right-hand side containing the matter distribution, which is supplied in the form of an equation of state (EOS) relating the pressure and the energy-density of the matter distribution.
	
	So far as the first technique is concerned, a large class of solutions are available in the literature, where, based on geometrically motivated functional forms of the metric potentials,  a large class of physically viable stellar models have been developed in the literature. Physical viability of such stellar solutions can be examined by analyzing the theoretical predictions with observational data. Tolman \cite{tolman} developed several such solutions by assuming the functional form of one of the metric potentials. Buchdahl \cite{buch} established limits on the compactness of a relativistic star. Amongst many others, Durgapal and Bannerji \cite{durga} and Finch and Skea \cite{finch} introduced polynomial and quadratic forms of the metric potentials to generate realistic stellar solutions. Of particular interest is the Vaidya and Tikekar \cite{vaidya} solution, which was constructed by identifying the fact that the $t =$ constant hypersurface of a spacetime, when embedded in a $4$-Euclidean space, has spheroidal rather than spherical geometry. Subsequently, the ansatz for the $g_{rr}$ metric potential was utilized to generate exact solutions capable of describing relativistic stars such as neutron stars. The procedure generates physically acceptable models for superdense stars without requiring an explicit equation of state. The Vaidya–Tikekar model has since then served as a foundation for numerous extensions involving anisotropy \cite{maurya19,maurya22,karmarkar0,paul,sharma}, charge \cite{koma,kumar,chatto,sharma2}, higher-dimensional effects \cite{khugaev,chanda}, as well as in modified gravity \cite{bhatta,kaur}.
	
	It should be stressed that the conventional way of developing a viable stellar model is to assume a specific form of the matter distribution vis-à-vis the EOS. In addition to neutron star models, for instance, Tooper \cite{tooper} analyzed polytropic stars in general relativity. Based on QCD arguments, Witten \cite{witten} proposed the strange matter hypothesis and subsequently, many authors have studied stellar properties by assuming the MIT bag model  EOS \cite{chodos,farhi,alcock}. Dey \textit{et al} \cite{dey} further refined these models with realistic density-dependent interactions for quark matter. More recent studies by Deb \textit{et al} \cite{deb}, Maurya \textit{et al} \cite{mau}, Patel \textit{et al} \cite{patel}, amongst many others, incorporate the effects of anisotropy and charge where linear, quadratic, polytropic, Chaplygin gas and colour-flavour-locked (CFL) EOS have been used.
	
	In the studies of relativistic stars, to generate exact solutions, an alternative method  is to impose geometric constraints on the system. One such constraint is the Karmarkar's condition \cite{karmarkar}. It is well known that in an embedding model, a four-dimensional Riemannian manifold can be characterized by the minimal number of extra flat dimensions needed to host it isometrically. This number is the embedding class of the manifold. A static and spherically symmetric spacetime of class I can be immersed in a flat five-dimensional Euclidean space, whereas a class II metric requires at least six dimensions. Karmarkar provided a necessary and sufficient condition for a static and spherically symmetric embedding of class II spacetime to be reducible to an embedding of class I, thus posing a geometric restriction on the system. This condition is derived based on the Riemann tensor components, which impose a relationship between the metric potentials $g_{tt}$ and $g_{rr}$, thereby reducing the number of independent functions and simplifying the solution-generating technique. Interestingly, many known solutions satisfying Karmarkar's condition yield physically viable stellar models \cite{bhar1,maur,bhar2,maur2,tello,ratan}.
	
	Another significant geometric condition in the studies of relativistic systems is the consideration of the vanishing of the Weyl tensor, which implies conformal flatness on the spacetime. The Weyl tensor is the trace-free component of the Riemann curvature tensor and it does not directly depend on the local matter but on the geometry itself. A vanishing Weyl tensor results in a highly symmetric spacetime where the local geometry is conformally equivalent to a flat spacetime. The simultaneous imposition of Karmarkar's condition and the vanishing Weyl tensor significantly narrows the class of admissible solutions leading to highly symmetric configurations.
	
	It is noteworthy that both these geometric conditions have been widely studied in the past either independently or with additional assumptions. Recently, Ratanpal \textit{et al} \cite{ratanpal} considered the simultaneous imposition of the Karmarkar's condition, conformal flatness and a vanishing complexity factor. The analysis shows that such an approach leads to  the Schwarzschild interior solution only. Earlier, Singh \textit{et al} \cite{singh} introduced pressure anisotropy $\Delta (r) = p_t - p_r,$  ($p_t =$ tangential pressure and $p_r =$ radial pressure) in class-I spacetimes and showed that in the isotropic limit $\Delta (r) = 0$, three possibilities emerge: a flat solution, the Schwarzschild interior solution and the Kohler-Chao solution. By imposing the pressure isotropy condition, the analysis was performed for a particular matter distribution. Such a treatment does not lead to all possible conformally flat configurations viz., the de Sitter geometry.
	
	It is important to note that isotropy and vanishing of the Weyl tensor are fundamentally distinct conditions. The isotropy condition ($\Delta(r) = 0$) arises from the equality of radial and tangential stresses in the matter distribution. It is a physical constraint on the energy-momentum tensor. The conformal flatness vis-a-vis vanishing Weyl condition ($C_{abcd}=0$) is a purely geometric condition. A spacetime may satisfy one without the other: an isotropic fluid need not necessarily generate a conformally flat geometry and a conformally flat geometry may, in principle, also correspond to an anisotropic source. This distinction naturally poses the following question: what are the possible geometries that simultaneously satisfy the Karmarkar's condition and the conformal flatness, without any prior matter assumption? Addressing this question is important for understanding how geometric constraints alone restrict the curvature of class-I spacetimes and whether the simultaneous imposition of these two curvature relations leaves any room for configurations possessing anisotropy or no-zero complexity factor. The present work is devoted to investigating this purely geometric connections in detail. It is worth noting that, although the resultant spacetimes are well known, a direct geometric derivation of the spacetimes from the simultaneous application of the Karmarkar's condition and conformal flatness condition has not been shown before. This derivation and its interpretation form the central motivation of the current investigation.
	
	The paper is organized as follows: In section \ref{sec2}, we lay down the equations resulting from the Karmarkar's condition together with conformal flatness condition for our assumed spacetime describing a spherically symmetric and static relativistic system. In section \ref{sec3}, making use of the above conditions, we show that only two distinct class of solutions can be generated by this method. Some concluding remarks are made in section \ref{sec4}.
	
	\section{Karmarkar's condition and conformal flatness}\label{sec2}
	We consider a spherically symmetric and static spacetime in the standard form as
	\begin{equation}
		ds_{-}^2 = e^{\nu(r)} dt^2 - e^{\lambda(r)}dr^2 - r^2(d\theta^2 + \sin^2\theta d\phi^2),\label{intm1}
	\end{equation}
	where $\nu(r)$ and $\lambda(r)$ are the undetermined functions. From the geometric point of view, to determine the two unknown metric potentials, we impose two independent tensorial constraints namely, the Karmarkar's (embedding-class I) condition and conformal flatness condition. 
	
	Note that the spacetime metric under consideration can be, in general,  associated with an embedding class II geometry. However, Karmarkar \cite{karmarkar} showed that the line-element can be transformed to an embedding class I geometry by imposing a particular condition. The condition ensures that a four-dimensional spacetime can be isometrically embedded into a five-dimensional flat Euclidean space, provided the following geometric constraint holds:
	\begin{equation}
		R_{1414} =
		\frac{R_{1212}R_{3434} + R_{1224}R_{1334}}
		{R_{2323}}, \label{kc1}  
	\end{equation}
	where $R_{2323} \ne 0$. For our assumed spacetime metric (\ref{intm1}), the Riemann tensor components are obtained as
	\begin{eqnarray}
		R_{1414} &=& -e^\nu \left(\frac{\nu^{''}}{2}+\frac{{\nu^{'}}^2}{4} -\frac{\lambda^{'}\nu^{'}}{4}\right), \label{r1}\\
		R_{2323} &=& -e^{-\lambda} r^2 \sin^2 \theta (e^{\lambda}-1) ,\label{r2}\\
		R_{1212} &=& -\frac{r \lambda^{'}}{2},\label{r3}\\
		R_{3434} &=& -\frac{1}{2} r \sin^2\theta \nu^{'} e^{\nu-\lambda},\label{r4}\\
		R_{1224} &=& 0, \label{r5}\\
		R_{1334} &=& 0.\label{r6}
	\end{eqnarray}
	Substituting Eqs.~(\ref{r1})-(\ref{r6}) in Eq.~(\ref{kc1}), we obtain
	\begin{equation}
		2\nu^{''}+{\nu^{'}}^2 = \frac{\nu^{'} \lambda^{'} e^\lambda}{e^\lambda-1}.\label{kc2}
	\end{equation}
	At this stage, we introduce the following transformations
	\begin{eqnarray}
		e^{-\lambda}=z,\label{z}\\
		e^{\nu/2}=y\label{y},
	\end{eqnarray}
	which allows us to write Eq.~(\ref{kc2}) in the form
	\begin{eqnarray}
		\frac{y^{''}}{y^{'}} &=& -\frac{z^{'}}{2z(1-z)},\label{e12}\\
		\Rightarrow \frac{dy^{'}}{y^{'}} &=& -\frac{dz}{2z(1-z)},\\
		\Rightarrow  y^{'} &=& A \sqrt{\frac{1-z}{z}},\label{yp}\\
		\Rightarrow y &=& A \int  \sqrt{\frac{1-z}{z}} dr +B, \label{yy}\\
		\Rightarrow e^\nu &=& \left[A \int  \sqrt{e^\lambda-1} dr +B\right]^2,\label{enu}
	\end{eqnarray}
	where $A$ and $B$ are integration constants.
	
	The conformal flatness of the spacetime can be enforced by the vanishing of the Weyl tensor $C_{abcd}$ (where, $a=r$, $b=\theta$, $c=\phi$ and $d=t$). For our assumed line-element, we have the following independent components
	\begin{eqnarray}
		C_{2121} &=& \frac{1}{24} \left(r^2 \lambda' \nu'-2 r \lambda'+4 e^{\lambda}-2 r^2 \nu''-r^2 \nu'^2+2 r \nu'-4\right),\label{c2121}\\
		C_{3131} &=& \sin^2 \theta C_{2121},\label{c3131}\\
		C_{3232} &=& -2 r^2 e^{-\lambda} \sin ^2\theta C_{2121},\label{c3232}\\
		C_{4141} &=& \frac{2 e^{n(r)}}{r^2}C_{2121},\label{c4141}\\
		C_{4242} &=& -e^{\nu-\lambda}C_{2121},\label{4242}\\
		C_{4343} &=& \sin ^2 \theta \left(-e^{\nu-\lambda}\right)C_{2121}.\label{4343}
	\end{eqnarray}
	Imposition of the condition of conformal flatness necessitates that all components of the Weyl tensor must vanish. We note that this requirement is fulfilled if we have
	\begin{eqnarray}
		r^2 \lambda' \nu'-2 r \lambda'+4 e^{\lambda}-2 r^2 \nu''-r^2 \nu'^2+2 r \nu'-4 = 0,\\
		\Rightarrow \frac{1-e^\lambda}{r^2}-\frac{\nu'\lambda'}{4}-\frac{\nu'-\lambda'}{2r}+\frac{\nu^"}{2}+\frac{\nu'^2}{4} = 0, \label{wyl}
	\end{eqnarray}
	which is same as the condition for conformal flatness in $D \ge 4$ dimensional spacetime derived earlier by Ponce de Leon \cite{leon}.
	
	\section {Exact conformally flat solutions from Karmarkar's condition}\label{sec3}
	Utilizing the transformations (\ref{z}) and (\ref{y}), we write the condition (\ref{wyl}) as
	\begin{equation}
		\frac{z-1}{zr^2}+\frac{y'z'}{2yz}-\frac{y'}{yr}-\frac{z'}{2zr}+\frac{y''}{y} = 0,\label{wylyz}   
	\end{equation}
	which, upon rearrangement, takes the form
	\begin{equation}
		rz'-2(z-1)+\left(2zr\frac{y'}{y}-r^2z'\frac{y'}{y}-2zr^2\frac{y''}{y}\right) = 0.
		\label{wyl2}
	\end{equation}
	
	Further, using Eq.~(\ref{e12}), we have
	\begin{equation}
		\Rightarrow \frac{y''}{y} = \frac{y''}{y'}\frac{y'}{y} = -\frac{z'}{2z(1-z)} \frac{y'}{y}.\label{ydp}
	\end{equation}	
	Inserting Eq.~(\ref{ydp}) in Eq.~(\ref{wyl2}), we obtain
	\begin{equation}
		\left(\frac{rz'}{z-1}-2\right)\left(z-1-\frac{y'zr}{y}\right) = 0.\label{factor}
	\end{equation}
	It is to be noted that $z = 1$ (i.e.\ $e^{\lambda} = 1$ and $e^\nu=$ constant) makes the denominator in Eq.~(\ref{factor}) vanish. 
	This corresponds to a flat spacetime with no gravitational field and is therefore physically trivial for self-gravitating configurations. Hence, we exclude the case $z = 1$ from our consideration. Physically meaningful solutions arise only for $z \neq 1$ as discussed below:

	\subsection{Solution I}
	One class of solution can be obtained from the relation
	\begin{eqnarray}
		\frac{rz'}{z-1}-2 = 0,\\
		\Rightarrow \frac{dz}{z-1} = \frac{2 dr}{r},\\
		\Rightarrow \ln (z-1) = 2\ln r+\ln K,\\
		\Rightarrow z=1+Kr^2,\\
		\Rightarrow e^\lambda = \frac{1}{1+Kr^2}. \label{el}
	\end{eqnarray}
	Substituting (\ref{el}) in (\ref{enu}), we have 
	\begin{equation}
		e^\nu = \left[\frac{A}{\sqrt{-K}}\sqrt{1+Kr^2}+B\right]^2, \label{enuf}
	\end{equation}
	where $K$ is the integration constant. Subsequently, the metric spacetime (\ref{intm1}) takes the form 
	\begin{equation}
		ds_{-}^2 = \left[\frac{A}{\sqrt{-K}}\sqrt{1+Kr^2}+B\right]^2 dt^2 - \frac{1}{1+Kr^2}dr^2 - r^2(d\theta^2 + \sin^2\theta d\phi^2).\label{metric}
	\end{equation}
	By redefine the constants as $K=-\frac{1}{R^2}$ and $-AR=D$, we obtain 
	\begin{equation}
		ds_{-}^2 = \left[B-D\sqrt{1-\frac{r^2}{R^2}}\right]^2 dt^2 - \frac{1}{1-\frac{r^2}{R^2}}dr^2 - r^2(d\theta^2 + \sin^2\theta d\phi^2),\label{sch}
	\end{equation}
	which is the well-known Schwarzschild interior solution \cite{sch} describing the interior of a homogeneous fluid sphere. 
	
	\subsection{Solution II}
	The second class of solutions can be obtained by solving the equation 
	\begin{equation}
		z-1-\frac{y'zr}{y} = 0.\label{sol2}
	\end{equation}
	Now, from Eq.~(\ref{yp}), we have
	\begin{equation}
		1-z^{-1} = -\frac{y'^2}{A^2}. \label{zz}
	\end{equation}
	Combining Eqs.~(\ref{sol2}) and (\ref{zz}), we obtain
	\begin{eqnarray}
		yy' = -A^2r,\\
		\Rightarrow y dy = -A^2r dr,\\
		\Rightarrow  y = P\sqrt{(1-\frac{r^2}{L^2})},\\
		\Rightarrow e^\nu = P^2(1-\frac{r^2}{L^2}),
	\end{eqnarray}
	where $P$ and $L=\frac{P}{A}$ are constants. Eq.~(\ref{zz}) yields
	\begin{eqnarray}
		z = 1-\frac{r^2}{L^2},\\
		\Rightarrow e^\lambda = \frac{1}{1-\frac{r^2}{L^2}}.
	\end{eqnarray}
	Thus, the line element (\ref{intm1}) takes the form
	\begin{equation}
		ds_{-}^2 = P^2\left(1-\frac{r^2}{L^2}\right) dt^2 - \frac{1}{1-\frac{r^2}{L^2}}dr^2 - r^2(d\theta^2 + \sin^2\theta d\phi^2),\label{desitt}  
	\end{equation}
	which is the de Sitter solution \cite{de}. 
	
	It is interesting to note that both the solutions obtained through this route correspond to idealized isotropic spacetimes. It is easy to show that the pressure anisotropy and the complexity factor vanish for such configurations. It should be pointed out here that, in the context of stability of text of  a star, Herrera \cite{Herrera} introduced the concept of complexity factor in terms of density inhomogeneity and anisotropic pressure of the star.  
	When the Karmarkar's condition and the conformal flatness are imposed, the geometry admits only isotropic distributions with zero complexity factor. 
	
	\section{Concluding remarks}\label{sec4}
	The simultaneous application of the Karmarkar's condition and the vanishing of the Weyl tensor provides a self contained geometric framework for studying relativistic systems. These two curvature relations act directly on the geometry of spacetime rather than on the matter variables and, when imposed together, they completely determine the metric potentials thereby providing highly restrictive physical matter compositions $T_{ab}~ (\rho, p)$. Further, in this formulation, both anisotropy and complexity factor vanish identically. 
	
	Our analysis highlights an important conceptual distinction between vanishing anisotropy and vanishing Weyl tensor. The condition $\Delta = p_t - p_r = 0$ arises from the equality of the fluid pressures and is therefore a matter constraint imposed on the energy-momentum tensor $T_{ab}$. In contrast, the condition $C_{abcd} = 0$ is geometric, ensuring conformal flatness. 
	A spacetime may satisfy one without the other: isotropy does not imply conformal flatness and conformal flatness does not necessarily imply isotropy. 
	
	We would like to point out that in a recent work of Singh \textit{et al} \cite{singh}, a similar analysis was carried out by solving the Einstein field equations for an anisotropic stellar configuration. In the limit $\Delta = 0$, the authors noted three possible cases: a flat spacetime metric, the Schwarzschild interior solution and the Kohler-Chao solution \cite{Kohler}. However, the Kohler-Chao spacetime is not conformally flat \cite{bharc} as its Weyl tensor does not vanish. Moreover, the technique  does not provide the de Sitter solution. This can be understood in terms of the approach adopted in Singh \textit{et al}'s paper. As the isotropy condition was applied to a static fluid distribution with positive pressure and density, suitable for describing the interior of a compact star, such a configuration must obey the matching conditions at a finite boundary and, therefore, should exclude a vacuum dominated equation of state of the form $p = -\rho$. Our approach is purely geometric and places no restriction on the physical source and consequently, the de Sitter  solution naturally appears as an admissible conformally flat, embedding class-I solution. In our formulation, the Karmarkar's condition and the vanishing Weyl tensor are imposed simultaneously and the isotropy of pressure and vanishing of the complexity factor are not assumed a priori; rather they emerge from the geometry itself.  
	
	In a recent pre-print, Ratanpal \textit{et al} \cite{ratanpal} combined the Karmarkar's condition, conformal flatness and an explicit assumption of vanishing complexity and showed that such assumptions lead to the Schwarzschild interior solution only. In this approach, the complexity condition serves as an additional physical constraint. Our calculation shows that the imposition of the  Karmarkar's condition together with the vanishing Weyl condition will only lead to an isotropic system with zero complexity factor. 
	
	Our analysis shows that if the geometry is restricted either by the Karmarkar's condition or the conformal flatness condition, it is possible to obtain new class of solutions only by incorporating other degrees of freedom. The current investigation shows an interplay between isotropy, conformal flatness and the embedding condition. Within the framework of general relativity, the combination of the Karmarkar's condition and the vanishing Weyl constraint leaves no degrees of freedom thereby yielding a highly constrained matter distribution. 

Obviously, to generate physically meaningful and interesting astrophysical models, conformal flatness condition must be relaxed if one applies the Karmarkar's condition. It is noteworthy that in a recent work, making use of the Karmarkar's condition, Maharaj \textit{et al}~\cite{maharaj2023} carried out a Lie group analysis to develop and study dynamically evolving systems. The analysis shows that, in general, there exist a $15-$dimensional Lie algebra of symmetries for the Karmarkar's condition in a conformally flat metric. Consequently, together with the Karmarkar's condition, it will be interesting to examine the Lie group analysis to generate physically meaningful new class of exact solutions for a given dynamical system.

\begin{acknowledgments}
RS gratefully acknowledges support from the Inter-University Centre for Astronomy and Astrophysics (IUCAA), Pune, India, under its Visiting Research Associateship Programme.
\end{acknowledgments}

\end{document}